\documentstyle[12pt]{article}

\input epsf

   \def\unlock{\catcode`@=11}

   \unlock

   \def\gsim{\mathrel{\mathpalette\@versim>}}

   \def\@versim#1#2{\vcenter{\offinterlineskip
        \ialign{$\m@th#1\hfil##\hfil$\crcr#2\crcr\sim\crcr } }}

\begin{document}
\renewcommand{\thefootnote}{\alph{footnote}}

\begin{titlepage}

\hspace*{\fill}\parbox[t]{2.8cm}{DESY 94-163 \\ SCIPP 94/27 \\ October 1994}

\vspace*{1cm}

\begin{center}
\large\bf
Transverse Momentum Distributions in Large-rapidity Dijet Production
at the Tevatron\footnote{Invited talk presented by V.D.D. at the
``${\rm VI}^{th}$ Rencontres de Blois'', Chateau de Blois, France, June 20-25,
1994}
\end{center}

\vspace*{0.5cm}

\begin{center}
Vittorio Del Duca \\
Deutsches Elektronen-Synchrotron \\
DESY, D-22607 Hamburg , GERMANY\\
\vspace*{0.5cm}
and\\
\vspace*{0.5cm}
Carl R. Schmidt \footnote{Supported in part by the U.S.
Department of Energy.} \\
Santa Cruz Institute for Particle Physics\\
University of California, Santa Cruz, CA 95064, USA
\end{center}

\vspace*{1.5cm}

\begin{center}
\bf Abstract
\end{center}

\noindent
In this contribution we examine the transverse momentum
distributions in dijet production at large rapidity intervals at the Tevatron,
using the BFKL resummation.
\end{titlepage}

\baselineskip=0.8cm

The state-of-the-art in jet physics at hadron colliders is described by
next-to-leading-order (NLO) QCD parton-level calculations\cite{GGK}.
They appear to be in very good
agreement with the one- and two-jet inclusive distributions obtained
from the data of the CDF experiment at the Fermilab Tevatron Collider.
However, since data are being collected at the CDF and D0 detectors at
larger and larger rapidities, it is possible to imagine kinematic
configurations where this fixed-order analysis is inadequate.
This could occur when the cross section contains large logarithms
 of the size of the rapidity interval in the scattering process.
If the initial parton momentum fractions are large,
then these logarithms factorize into the partonic subprocess cross
section and can be resummed by using the techniques of Balitsky,
Fadin, Kuraev, and Lipatov (BFKL)\cite{BFKL}.

In analyzing dijet production
experimentally so that it most closely resembles the configuration
assumed in the BFKL theory, the jets are ordered first by their
rapidity rather than by their energy\cite{MN}.
Thus, we look at all the jets in the event
that are above a transverse momentum cutoff $p_{\perp min}$,
using some jet-definition algorithm, and rank them by their rapidity.
We then tag the two jets with the largest and smallest rapidity
$(\vec{p}_{1\perp}, y_1)$ and $(\vec{p}_{2\perp},y_2)$, where
we always take $y_1>y_2$, and
observe the distributions as a function of these two {\it tagging jets}.
We reexpress the jet rapidities in terms of the
rapidity interval $y=y_1-y_2$ and the rapidity boost ${\bar y}=(y_1+y_2)/2$.
This is convenient since we are mainly interested in the behavior of the
parton subprocess, which does not depend on $\bar y$. Then we
sum inclusively the hadrons or jets produced in the rapidity interval $y$
between the tagging jets, and refer to them as {\it minijets}.
For large values of $y$ the cross section for this process can be written
\begin{equation}
{d\sigma_0\over dy\,d{\bar y}\,dp_{1\perp}^2 dp_{2\perp}^2 d\phi}\,=\,
x^0_Ax^0_B\,f_{\rm eff}(x^0_A,\mu^2)f_{\rm eff}(x^0_B,\mu^2)\,
{d\hat\sigma_{gg}\over dp_{1\perp}^2 dp_{2\perp}^2 d\phi}\ ,
\label{general}
\end{equation}
where the parton momentum fractions are dominated by the contribution
from the two tagging jets
\begin{eqnarray}
x^0_A &=& {p_{1\perp} e^{y_1} \over\sqrt{s}}\nonumber\\
x^0_B &=& {p_{2\perp} e^{-y_2} \over\sqrt{s}},\label{pmf}
\end{eqnarray}
$\phi$ is the azimuthal angle in the transverse plane
and $\mu$ is the factorization/renormalization scale.
In this limit the amplitude is dominated by $gg$, $qg$, and $qq$
scattering diagrams with gluon-exchange in the $t$-channel.  The
relative magnitude of the different subprocesses is fixed by the color
strength of the respective jet-production vertices, so it
suffices to consider only $gg$ scattering and to include the other
subprocesses by means of the effective parton distribution
function $f_{\rm eff}(x,\mu^2)$\cite{MN}.

The higher-order corrections to the $gg$ subprocess cross section in
(\ref{general}) can be expressed via the solution of the BFKL
equation\cite{BFKL}, which is an all-order resummation in
$\alpha_s$ of the leading powers of the rapidity interval
\begin{equation}
{d\hat\sigma_{gg}\over dp_{1\perp}^2 dp_{2\perp}^2 d\phi}\,=\,
{C_A^2\alpha_s^2 \over 4\pi p_{1\perp}^3 \, p_{2\perp}^3} \,
\sum_n e^{in(\phi-\pi)} \int_0^{\infty} d\nu e^{\omega(n,\nu)\, y}
\cos\left(\nu \, \ln{p_{1\perp}^2 \over p_{2\perp}^2} \right)
\label{mini}
\end{equation}
with
\begin{equation}
\omega(n,\nu) = {2 C_A \alpha_s \over\pi} \bigl[ \psi(1) -
 {\rm Re}\,\psi ({|n|+1\over 2} +i\nu) \bigr],
\label{eigen}
\end{equation}
and $\psi$ the logarithmic derivative of the Gamma function.

In ref.~\cite{DDS}
we found that at Tevatron energies the transverse momentum
$p_\perp$ distribution and the jet-jet correlations in $p_\perp$ and
$\phi$ are significantly affected by the minijet resummation.
For instance, the $p_\perp$ distribution was considerably enhanced at large
$p_\perp$ and large $y$.
However, in comparing the truncation of the BFKL resummation (\ref{mini})
to ${\cal O}(\alpha_s^3)$ with the exact
${\cal O}(\alpha_s^3)$ calculation of dijet production,
computed through the 2$\rightarrow$3 parton amplitudes\cite{DDS2},
we noticed that the large-rapidity approximation to the kinematics
seriously overestimates the cross section and causes a serious error in
the BFKL predictions when the two tagging
jets are not back-to-back in $p_{\perp}$ and $\phi$, even for rapidity
intervals as large as $y=6$.
This occurs because the large-$y$ cross section assumes that
the third (minijet) parton can be produced anywhere within the rapidity
interval $[y_2,y_1]$ with equal probability, whereas in the full
$2\rightarrow3$ cross section the probability is highly suppressed by the
structure functions when the third jet strays too far from the center of this
interval. In order to account for this error we introduced in ref.\cite{DDS2}
an effective rapidity $\hat y$ to take into account the fact
that the range in rapidity spanned by the minijets is
typically less than the kinematic rapidity interval $y$. $\hat y$
is defined so that if we replace $y\rightarrow\hat y$ in the BFKL solution
the difference $y\!-\!\hat y$ is nonleading. Since the rapidity
variable which is resummed by BFKL is only defined up to transformations
$y\rightarrow y+X$ where $X$ is subleading at large rapidities, we used
$\hat y$ instead of $y$ in the BFKL resummation in order to obtain
quantitatively more reliable predictions of the transverse momentum
distributions. We found that the effects on the $p_\perp$ distribution are
not as dramatic as we had predicted in ref.\cite{DDS} using the kinematic
rapidity $y$. Because of the relatively small deviations of the BFKL
resummation with the effective
rapidity $\hat y$ from the Born-level calculation, and the sizeable
renormalization/factorization scale ambiguities in the BFKL approximation,
we concluded that a complete NLO calculation could probably give a more
reliable estimate to the $p_{\perp}$ distributions.  However, much of the
uncertainties due to the renormalization/factorization scale drop out in
the ratios of cross sections, so the ratio of $p_{\perp}$ distributions
should be a good observable to examine with the BFKL resummation.

In Fig.~1 we plot the ratio of transverse momentum distributions of jet 1
\begin{equation}
r(\mu^2) = {d\sigma(p_{2\perp min,a})/dyd\bar y dp_{1\perp} \over
d\sigma(p_{2\perp min,b})/dyd\bar y dp_{1\perp}}
\end{equation}
calculated using the effective rapidity $\hat y$ in the BFKL resummation,
with two different cutoffs for jet 2 transverse momentum,
$p_{2\perp min,a}=20$ GeV and $p_{2\perp min,b}=30$ GeV.
The rapidity boost $\bar y$ is integrated over, subject to the constraint
$|y_1|_{max}=|y_2|_{max}=3.2$. The rapidity interval is integrated in unit
bins centered around $y=4$ and $y=5$.
We use the LO CTEQ2 parton distribution functions\cite{cteq}
with two extreme choices for the ren./fact. scale
$\mu^2 = 4 max(p_{1\perp}^2, p_{2\perp}^2)$ for the lower curves and
$\mu^2 = p_{1\perp} p_{2\perp}/4$ for the upper ones.

\begin{figure}[htb]
\vspace{12pt}
\vskip-4cm
\epsfysize=16cm
\centerline{\epsffile{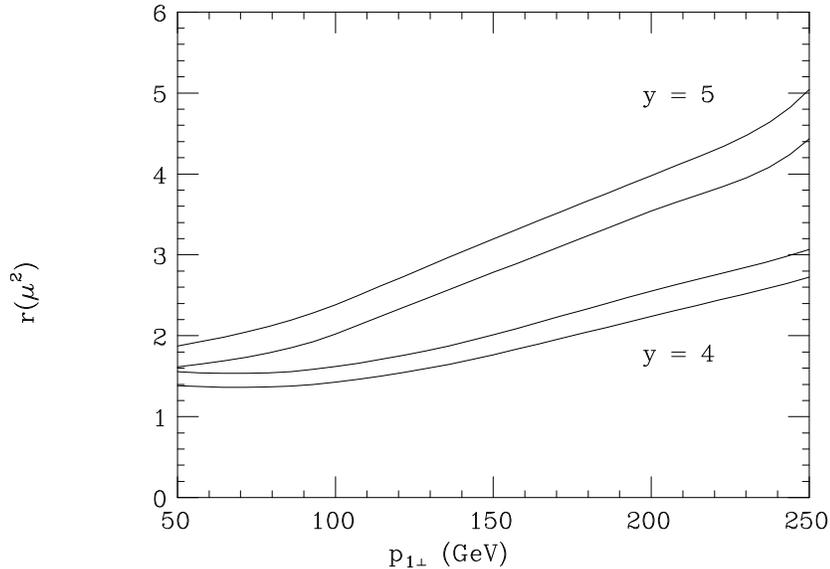}}
\vspace{18pt}
\vskip-5cm
\caption{The ratios of transverse momentum distributions of jet 1 at
$y=4$ and 5,
with two different cutoffs for jet 2 transverse momentum,
$p_{2\perp min,a}=20$ GeV in the numerator and $p_{2\perp min,b}=30$ GeV in
the denominator. The ren./fact. scale is set to $\mu^2 = 4 max(p_{1\perp}^2,
 p_{2\perp}^2)$ for the lower curves and
$\mu^2 = p_{1\perp} p_{2\perp}/4$ for the upper ones.}

\label{fig:ratios}
\vspace{12pt}
\end{figure}

{}From the plot we see that lowering the $p_\perp$ cutoff for the second
jet significantly increases the cross section, particularly for large
$y$ and $p_{1\perp}$.  For example, for $y=5$ and $p_{1\perp}=300$
GeV, we gain more than a factor of six in lowering $p_{2\perp min}$
from 30 GeV to 20 GeV.  This enhancement is entirely due to events in
which the tagging jets are very unbalanced in $p_\perp$, thus
requiring $\ge$ 3 final state jets.  This suggests that a beyond-NLO
calculation, such as the BFKL resummation, may be necessary for this
configuration.

\end{document}